\pdfoutput=1
\documentclass[conference]{IEEEtran}
\usepackage{amsmath}
\usepackage{amssymb}
\usepackage[usenames]{color}
\let\labelindent\relax
\usepackage{enumitem}
\usepackage{graphicx}
\usepackage{listings}
\usepackage{multirow}
\usepackage{relsize}
\usepackage{soul}  
\usepackage{fancybox}
\usepackage{subfig}
\usepackage{textcomp}
\usepackage{times}
\usepackage{tipa}
\usepackage{url}
\usepackage{algorithm}
\usepackage{algorithmicx}
\usepackage[noend]{algpseudocode}
\usepackage{algpseudocode}
\makeatletter
\renewcommand{\ALG@beginalgorithmic}{\small}
\makeatother
\usepackage{booktabs}
\usepackage{bigstrut}

\usepackage{lipsum}

\newcommand{\diver}{\textsc{Diver}}
\newcommand{\distea}{\textsc{DistEA}}
\newcommand{\tool}{\distea}
\newcommand{\tech}{\tool}

\newcommand{\EAS}{\textsc{EAS}}
\newcommand{\PI}{\textsc{PathImpact}}

\newcommand{\CI}{\textsc{CoverageImpact}}

\newcommand{\LTS}{\textsc{LTS}}

\def\denseitems{
  \itemsep1pt plus1pt minus1pt
  \parsep0pt plus0pt
  \parskip0pt\topsep0pt}

\usepackage{color}

\usepackage{listings,mdframed}
\usepackage{breqn}

\usepackage{pifont,calc}
\definecolor{verylightgray}{gray}{0.85}
\newlength\lsthorizontalpadding
\newcommand*\lstnumberstyle{\ttfamily\scriptsize}
\newlength\lstnumbersep
\newlength\lstnumberwidth
\usepackage[none]{hyphenat}

\usepackage[hidelinks]{hyperref}
\usepackage{cite}
\usepackage[table]{xcolor}

\usepackage{times}
\usepackage{textcomp}
\usepackage{balance}

\definecolor{Gray}{gray}{0.85}
\definecolor{LightCyan}{rgb}{0.88,1,1}

\newcolumntype{g}{>{\columncolor{Gray}}c}
\newcolumntype{w}{>{\columncolor{white}}c}

\definecolor{lightblue}{rgb}{.90,.95,1}

\newcommand{\haipengemail}{hcai@nd.edu}
\newcommand{\raulemail}{rsanteli@nd.edu}
\newcommand{\dougemail}{dthain@nd.edu}

\newfont{\ttlsc}{phvb8t at 15pt}
\newfont{\autt}{pcrr8t at 11pt}
\begin{document}
\pagenumbering{arabic}
  \pagestyle{plain}
\setlength{\parskip}{0pt} 

\title{{\distea}: Efficient Dynamic Impact Analysis\\for Distributed Systems\vspace{-1.5ex}}


\author{Haipeng Cai and Douglas Thain \\
University of Notre Dame, Indiana, USA \\
email: {\tt\{\href{mailto:\haipengemail}{hcai}|\href{mailto:\dougemail}{dthain}\}@nd.edu}
}

\maketitle
\thispagestyle{plain}


\begin{abstract}
Dynamic impact analysis is a fundamental technique for
understanding the impact of specific program entities, or changes to them, on the rest of
the program for concrete executions.
However, existing techniques are either inapplicable or of very limited utility for
distributed programs running in multiple concurrent processes.
This paper presents {\tech}, a technique and tool for dynamic impact analysis of
distributed systems.
By partially ordering distributed method-execution events and inferring causality from
the ordered events, {\tech} can predict impacts propagated both within and
across process boundaries.
We implemented {\tech} for Java and applied it to
four distributed programs of various types and sizes, including two enterprise systems.
We also evaluated the precision and practical usefulness of {\distea}, and demonstrated
its application in program comprehension, through two case studies.
The results show that {\distea} is highly scalable, more effective than
existing alternatives, and instrumental to understanding distributed systems and their executions.
\end{abstract}

\IEEEpeerreviewmaketitle


\section{Introduction}\label{sec:intro}
%

Program changes drive the evolution of software systems, yet also 
pose threats to their quality and reliability~\cite{rajlich2006changing}.
Thus, it is crucial to understand potential consequences of those changes
even \emph{before} applying them to candidate program locations.
To accomplish this task, developers need to perform impact
analysis~\cite{bohner96jun,rovegard2008empirical,tao2012software}
with respect to those locations, an integral step of modern software development process~\cite{Rajlich2014SEM}.
In particular, for developers working with specific operational profiles of the program, dynamic impact analysis~\cite{bohner96jun,Li2013ASC} is an attractive option
as it narrows down the search space of such impacts to the concrete context of those profiles.


During the past two decades, research on dynamic impact analysis has been extensively invested~\cite{Li2013ASC}, resulting in
a rich and diverse set of relevant techniques and tools (e.g.,~\cite{law03may,apiwattanapong05may,cai14diver,cai15diverplus}).
However, most of existing such approaches mainly address sequential programs only, with much less targeting
concurrent yet centralized software~\cite{park2010falcon,goswami2000dynamic,krinke2003context,xiao2005improved,giffhorn2009precise} while very few
applicable to distributed systems. 
On the other hand, to accommodate the increasingly demanding performance and scalability needs of today's computation
tasks, more distributed systems than centralized ones are being deployed, raising
an urgent call for technical supports, including impact analysis, for effective maintenance and evolution of
those systems~\cite{park2010falcon,jayaram2011program,garcia2013identifying,beschastnikh2014inferring}.

Code analysis techniques for distributed systems were 
explored since early on~\cite{korel1992dynamic,kamkar1995dynamic,duesterwald1993distributed}
and ratcheted up recently~\cite{mohapatra2006distributed,barpanda2011dynamic,pani2012slicing,octeau2013effective},
largely focusing on detailed analysis of program dependencies. 
However, the majority of these approaches were designed only for procedural programs~\cite{barpanda2011dynamic}.
For distributed object-oriented programs, \emph{backward} dynamic slicing algorithms have been developed, yet it is still
unclear whether they can work with real-world systems~\cite{mohapatra2006distributed,barpanda2011dynamic}.
And for impact analysis, \emph{forward} slicing would be needed.
Nevertheless, the fine-grained (statement-level) analysis used by
slicing would be overly heavyweight for impact analysis commonly
adopted at method level~\cite{bohner96jun,Li2013ASC,law03may,apiwattanapong05may,cai14diver}.

Unfortunately, 
developing an efficient dynamic impact analysis for distributed systems remains 
challenging.
One major difficulty lies in the lack of explicit invocations or references among decoupled components in such systems
~\cite{goswami2000dynamic,jayaram2011program,Tragatschnig2014IAE},
whereas
traditional approaches usually rely on those explicit information to compute dependencies for impact prediction.
Lately, various dependence-analysis techniques other than slicing have also been proposed~\cite{popescu2012impact,garcia2013identifying,Tragatschnig2014IAE}.
While efficient for \emph{static} impact analysis, these approaches are limited to 
systems of special type
such as distributed \emph{event-based} systems (DEBS)~\cite{muhl2006distributed}, or rely on specialized
language extensions like EventJava~\cite{eugster2009eventjava}.
Other approaches are potentially applicable in a wider scope, yet they depend on information not always available, such as execution logs of particular pattern~\cite{lou2010mining}, or suffer from overly-coarse granularity (e.g., class-level) ~\cite{popescu2012impact,lou2010mining,garcia2013identifying} and/or unsoundness~\cite{murphy1996lightweight}, in addition
to imprecision, of their analysis results.

In this paper, we present {\tech}, a dynamic impact analysis approach 
 for commonly deployed distributed systems where components 
communicate via socket-based message passing.\footnote{\vspace{-8pt}We distinguish components as such that each runs in a separate process.}
By exploiting 
the happens-before relations~\cite{lamport1978time} among distributed method-execution events, {\tech} predicts impacts of one method on others of a given system
both within and across its 
concurrent processes.
Akin to the execute-after-sequences ({\EAS}) technique~\cite{apiwattanapong05may},
our approach offers results that are safe relative to the concrete executions utilized 
with high efficiency, while relying on neither well-defined inter-component interfaces nor message-type specifications as
needed by peer approaches (for DEBS).

We evaluate {\tech} on four distributed Java programs, including two
enterprise systems, and demonstrate that it is able to work with large distributed systems
with both blocking and non-blocking (e.g., selector-based~\cite{artho2013software}) communications.
In the absence of peer techniques
directly comparable to ours, we take a coverage-based approach~\cite{orso03sep}, which
reports as impacted all methods covered in the utilized executions, as a safe baseline alternative, and
measure the effectiveness of {\distea} against it. 
The results show that {\tech} 
can greatly reduce
the size of potential impacts to be inspected,
by 36\% on average, 
relative to the baseline,
at the mean cost of 50 seconds to finish the one-time static analysis and three seconds to answer a query, with a
runtime overhead of 11\%.
Since there is no automatic approach available to us for computing ground-truth impacts either, we manually evaluate the
precision of {\distea} on five randomly selected cases in a case study. Also, we explore the usefulness of
{\distea} in program comprehension in a second case study. Our results suggest that developers using {\distea}
may expect an average precision of about 60\% and considerable benefits for understanding distributed programs
and executions. 

The main contributions of this work include:
\begin{itemize}[leftmargin=14pt]
\denseitems
\item A dynamic impact analysis, {\distea}, for 
    distributed systems where concurrent processes communicate via socket-based message passing
(Section~\ref{sec:tech}).
\item An implementation of {\distea} for Java working with large enterprise distributed systems 
with both blocking and non-blocking communications (Section~\ref{sec:impl}).
\item An empirical study of {\distea} 
    showing its promising effectiveness and scalability
    (Section~\ref{sec:eval}).
\item Two in-depth case studies of {\distea} showing its practical usefulness for impact analysis and 
 benefits for distributed-program understanding (Section~\ref{sec:appendix}).
\end{itemize}

\begin{figure}[t]
 \centering
 \lstset{basicstyle=\small}
  \begin{minipage}[t]{.48\textwidth}
  \begin{mdframed}
  \vspace{-10pt}
   \begin{lstlisting}
    public class C {
        Socket csock = null;
        public C(String host, int port) {
            csock = new Socket(host, port); }
        void shuffle(String s) {...}
        char compute(String s) {
            shuffle(s);
            csock.writeChars(s);
            return sock.readChar(); }
        public static int main(String[] a) {
            C c = new C('localhost',2345);
            System.out.println( c.compute(a[0]) );
            return 0; }}
    public class S {
        Socket ssock = null;
        public S(int port) {
            ssock = new Socket(port);
            ssock.accept(); }
        char getMax(String s) {...}
        void serve() {
            String s = ssock.readLine();
            char r = getMax(s);
            ssock.writeChar(r); }
        public static int main(String[] a) {
            S s = new S(2345);
            return s.serve(); }}
    \end{lstlisting}
    \vspace{-10pt}
 \end{mdframed}
\end{minipage}
 \vspace{-6pt}
 \caption{An example distributed program $E$ consisting of two components: C (client) and S (server).}
 \label{fig:csexample}
  \vspace{-20pt}
\end{figure}


\section{Motivation and Background}\label{sec:motbkg}
In this section, we first present a usage scenario of dynamic impact analysis that motivates
our development of {\distea}. Then, we give necessary background on techniques
underlying the design of {\distea}.

\subsection{Motivating Example}
When maintaining and evolving a distributed program which consists of multiple components,
the developer needs to understand potential change effects not only in the component where the change
is proposed, but also those in all other components.
%
To achieve better flexibility and scalability, these components are usually loosely coupled or entirely decoupled as a result of
implicit invocations among them realized via socket-based message passing, which, however, reduces
the utility of existing impact analysis to a very limited extent.

Consider the example program $E$ shown in Figure~\ref{fig:csexample}, which consists of
two components: a server and a client, implemented in classes {\tt S} and {\tt C}, respectively.
The client simply retrieves the largest character in a given string by sending the task to the server, which finishes the task and sends the result back to the client. Suppose now the developer proposes to apply a new algorithm in the {\tt S::getMax} method as part of an upgrade plan for the server and, thus, needs to determine which other parts of the program may have to be changed as well. Having an available set $I$ of inputs, the developer wants to perform a dynamic impact analysis to get a quick but safe estimation on potential impacts of the candidate change with respect to $I$.

At first glance, it seems that the developer has many options (e.g., ~\cite{apiwattanapong05may,cai14diver,cai15diverplus})
to accomplish this task. Unfortunately, it soon turns out that those existing options have
merely quite limited utility in this
context. Since there is no explicit dependencies between {\tt S} and {\tt C}, existing dynamic impact analysis would
predict impacts within the \emph{local} component (i.e., where the changes are located; $S$ in this case) only.
In consequence, the developer would have to ignore impacts in \emph{remote} components ($C$ in this case),
or make a worst-case assumption that all methods in remote components are to be impacted.

As illustrated by this example, the distributed system we address in this work is one in which
components located at networked computers communicate and coordinate their actions only by passing messages~\cite{Coulouris2011DSC}: The components run concurrently in multiple processes \emph{without a global clock}.
\subsection{Dynamic Impact Analysis}\label{subsec:eas}
Typically, a dynamic impact analysis technique inputs a program $P$, an input set $I$, and a
\emph{query set} $M$ (the set of methods for which impacts are to be queried),
and outputs an \emph{impact set} (the set of methods in $P$ potentially impacted) of $M$ when running $I$.
One representative such technique is based on the execute-after-sequences ({\EAS})~\cite{apiwattanapong05may}, which
computes impacts from the execution order of methods.
Given a query $c$, {\EAS} considers all methods that execute after
$c$ as potentially affected by $c$ or by any changes to it.

To find the method execution order, 
{\EAS} records two main method-execution events
using two integers for each method $m$:
the first time $m$ is entered
and the last time program control is returned into $m$. 
Then, the analysis infers the execute-after relations according to the occurrence time of those events. 
In presence of multi-threaded executions,
{\EAS} monitors also method returns and treats them as returned-into events.
For the concrete set of executions, no methods that never executed after the query $c$ can be impacted by it; thus,
the results produced by {\EAS} are safe (i.e., of 100\% recall) relative to those executions.
However, an execute-after relation does not always lead to an impact relation since a method may execute after the query
yet has no any dependence on that query; thus, {\EAS} is imprecise due to its conservative nature~\cite{cai15jss}.

On the other hand, the need for maintaining only little information (i.e., the two integers per method) enables the high
efficiency of {\EAS}. Therefore, despite of its known imprecision, impact analysis using execute-after relations like
{\EAS} remains a viable option, especially for users who desire getting a safe approximation of impacts quickly, such as
the developer in the above example scenario. In fact, to the best of our knowledge, {\EAS} is still the most
efficient dynamic impact analysis to this date~\cite{orso04may,cai14diver,cai15diverplus}.
Thus, as the first
attempt exploring efficient dynamic impact analysis for distributed systems, we start with an
{\EAS}-based approach in this work.



Nonetheless, {\EAS} itself does not work with distributed programs. 
One may attempt to first apply this technique to each component independently,
generate the execute-after sequences for each (process), and then compute
impacts for the entire system by referring to all the event sequences.
Unfortunately, different components often run asynchronously and on machines of different physical clocks~\cite{beschastnikh2014inferring,lou2010mining}, easily leading to incorrect order of
method executions with respect to all processes, hence erroneous impact sets. 


\subsection{Timing in Distributed Systems}\label{subsec:lts}
Different approaches exist to manage the timing in distributed systems
~\cite{lamport1978time,fidge1988timestamps,mattern1989virtual}, of which the one
by Lamport~\cite{lamport1978time} maintains a logical clock per process to partially order distributed events
over all processes with a simple algorithm,
which we refer to as the \emph{Lamport TimeStamping} ({\LTS}) algorithm.
%
The {\LTS} approach first defines
a logical clock $C_i$ for each process $P_i$, which is a function that assigns a number $C_i\langle$$a$$\rangle$ to an event $a$ in $P_i$.
Based on this definition, an event $a$ happened before another event $b$ if the number assigned to $a$ is less than that
assigned to $b$, or formally
\vspace{-3pt}
\begin{equation}
\vspace{-3pt}
a\longrightarrow~b~\Longrightarrow~C\langle a\rangle~<C\langle b\rangle
\end{equation}
which is called the \emph{clock condition}. Then, to maintain the clock condition during system executions, the following
rules~\cite{lamport1978time} should be observed by each process:
\begin{itemize}[leftmargin=14pt]
\denseitems
\item Each process $P_i$ increments $C_i$ between any two successive events that happened in $P_i$.
\item If event $a$ is that process $P_i$ sends a message $m$, then the message contains a timestamp $T_m$=$C_i\langle a\rangle$.
\item When a process $P_j$ receives a message $m$, it sets $C_j$ greater than or equal to its current value and greater than $T_m$.
\end{itemize}

%

For our {\EAS}-based approach to impact analysis for distributed systems, the {\LTS} algorithm can be
utilized to preserve the partial ordering of 
distributed events across multiple processes running on separated machines. Furthermore, this ordering 
would enable inferring
causality between methods 
hence the computation of 
impacts of one method on others 
both within and across system components. 




\section{Approach}
\label{sec:tech}
To achieve an efficient dynamic impact analysis for distributed programs, 
{\distea} utilizes only lightweight runtime information on method execution order.
%
We first 
present the fundamentals underlying our approach, including
the definition of method events used by {\distea} and its rationale for impact prediction.
Then, we give an overview and illustration on the inner workings of {\distea} 
followed by details on the analysis algorithms.

\subsection{Fundamentals}
\subsubsection{Method-Execution Events}
In the general context of distributed systems, an event is defined as any happening of interest observable from
within a computer~\cite{lamport1978time}. More specifically, 
events in a DEBS are often expressed as messages transferred among system components and defined by a set of attributes~\cite{muhl2006distributed,garcia2013identifying}.
In contrast, while it also deals with message passing in distributed systems, {\distea} neither makes
any assumption nor reasons about the structure or content of the messages. Particularly, for dynamic impact analysis, {\distea}
%
monitors and utilizes two major classes of events as defined below:
\begin{itemize}[leftmargin=14pt]
\item{\textbf{Communication Event.}} A communication event $E_C$ is an occurrence of message transfer between two components $c1$ and $c2$, denoted as $E_C$($c1$,$c2$) if $c1$ initiates $E_C$ which attempts to reach $c2$. Further, according to the direction of message flow, we distinguish two major subcategories of such events: \emph{sending a message} to a component and \emph{receiving a message} from a component.
\item{\textbf{Internal Event.}} An internal event $E_I$ is an occurrence of method execution within a component $c$, denoted as $E_I$($c$). 
    Further, we differentiate three subcategories of internal events: 
    \emph{entering a method}, \emph{returning from a method}, and \emph{returning into a method}, denoted as $m_e$, $m_x$, and $m_i$, respectively, for the relevant method $m$.
\end{itemize}

For internal events, we capture both the return 
and returned-into 
 events for each method. 
However, we distinguish them during static analysis 
only and treat them equally in the 
monitoring algorithm (Section~\ref{subsec:easalgo}).
The reason is to correctly identify execute-after relations from interleaving method executions in
multiple threads 
as detailed in~\cite{apiwattanapong05may}.


\subsubsection{Impact Inference}
One challenge to developing {\distea} is to infer the execute-after relations in the presence of asynchronous events over multiprocess concurrent executions. Fortunately, 
maintaining a logical notion of time per process to discover just a partial ordering of 
method-execution events suffices for that inference required in {\distea}. 

In essence, the execute-after (EA) relation between any two methods can be semantically
deduced from the happens-before relation between relevant internal events of corresponding methods;
and the partial ordering of the internal events reveals such happens-before relations~\cite{lamport1978time}. 
Formally, given two executed methods $m1$ and $m2$, we have
%
%
%
\vspace{-3pt}
\begin{equation}\label{eq_ea}
\vspace{-3pt}
m1_e \prec m2_x \bigvee m1_e \prec m2_i \Longrightarrow EA(m2,m1)
\end{equation}
where $\prec$ is the \emph{happens-before} relation.
Without loss of generality, $m1_e$$\prec$$m2_x$ and $m1_e$$\prec$$m2_i$ both imply that ``$m2$ executes after $m1$, thus $m2$ may be affected by $m1$ or by any changes to $m1$", hence the execute-after relation \emph{EA}($m2$,$m1$) between $m1$ and $m2$.

Based on the above inference, for a given query $c$, computing the impact set $IS$($c$) of $c$ 
is reduced to retrieving methods, from multiprocess method-execution event sequences, that satisfy the partial ordering of
internal events of candidate methods as follows:
\vspace{-3pt}
\begin{equation}\label{eq_is}
\vspace{-3pt}
IS(c)~=~\{m\mid c_e\prec m_i \vee c_e \prec m_x\}
\end{equation}
Note that only 
internal events are directly used for impact inference, while
communication events 
are utilized to maintain the partial ordering of internal events in all processes.

\subsection{{\distea} Overview}
\subsubsection{Process}\label{subsec:processflow}
\begin{figure}[t]
  \begin{center}
  \includegraphics[scale=0.7]{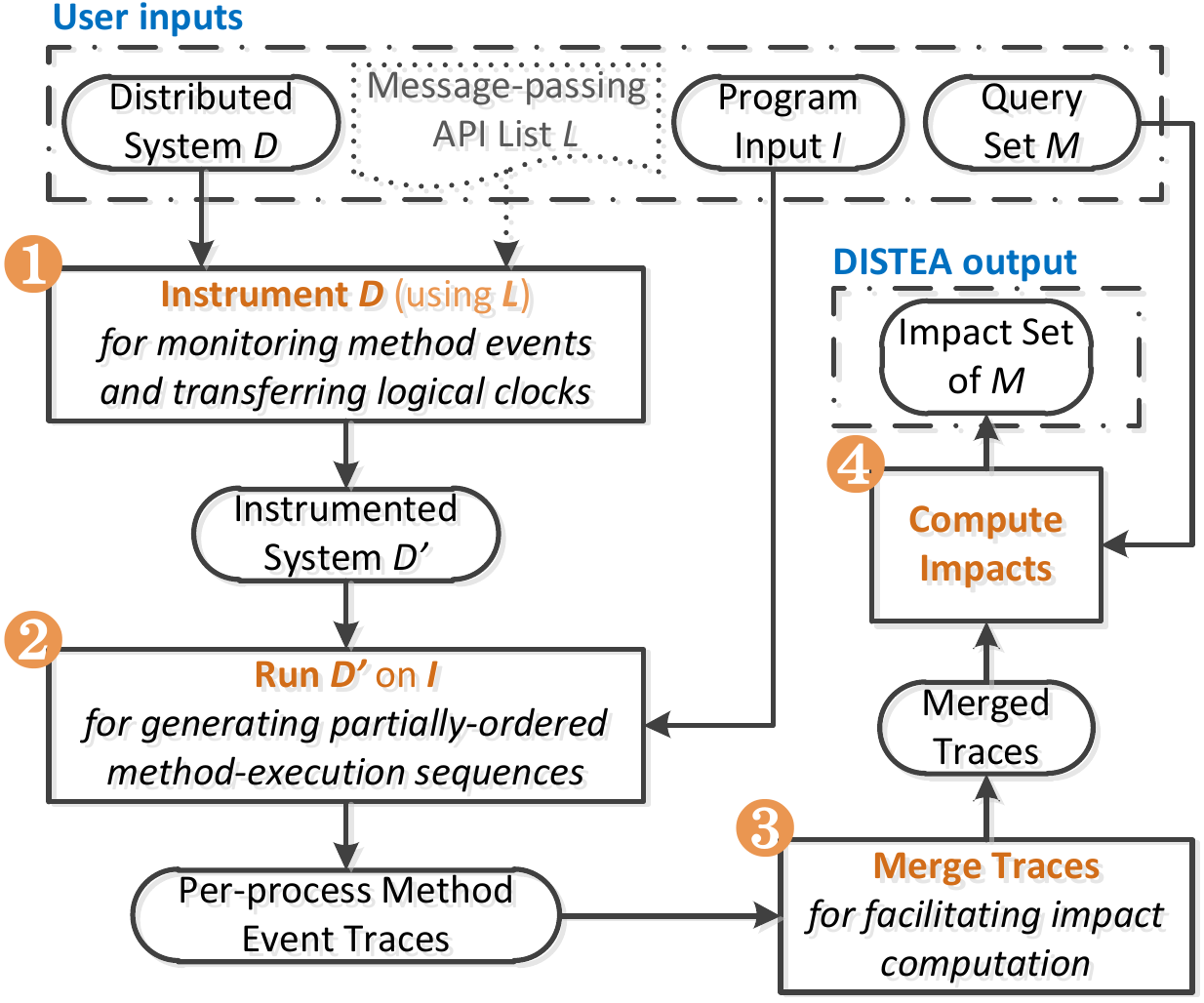}
  \vspace{-6pt}
  \caption{The overall process flow of {\distea}, where the numbered steps are detailed in Section~\ref{subsec:processflow}.}
  \label{fig:disteaprocess}
  \end{center}
  \vspace{-25pt}
\end{figure}

The overall process flow of our technique is depicted in Figure~\ref{fig:disteaprocess}, where
the three primary inputs are the system $D$ under analysis,
a set $I$ of program inputs for $D$, and a query set $M$.
An optional input, a message-passing API list $L$ can also be specified to
help {\distea} identify program locations where probes for communication events
should be instrumented (as detailed in Section~\ref{sec:impl}). The output of {\distea} is a set of potential impacts
of $M$ computed from the given inputs in four steps as annotated in the figure.

The \emph{first step} performs the static analysis in {\distea}, where the input program $D$ is instrumented for both monitoring
method-execution events and synchronizing logical clocks among concurrent processes, and
the instrumented version $D'$ of $D$ is produced.
Then, the \emph{second step} executes $D'$ on the given input set $I$, during which internal events are produced
and time-stamped by means of communication events such that the partial ordering for all internal events is preserved. 
Next, in the \emph{third step}, method event traces generated from all processes are gathered and
merged to a holistic ordered sequence stored in either one or multiple traces. 
Finally, the \emph{fourth step} takes the query set $M$ and the merged event sequence
to compute the impact set of $M$. 

\subsubsection{Illustration}\label{subsec:illustrate}
To illustrate the above process flow, consider the example program $E$ of Figure~\ref{fig:csexample}.
{\distea} first instruments $E$ and produces instrumented code for both components. Next, suppose the instrumented
server and client components $S'$ and $C'$ are deployed on two distributed machines, 
and $S'$
starts first before an user launches $C'$. 
When running concurrently, $S'$ and $C'$ generate two method-event sequences in two separate processes,
as listed \emph{in full}
in the first and last two columns in Table~\ref{tab:examplelts}, respectively.
As is shown, logical clocks are updated upon communication events. For instance, the logical clock of the server
process is first updated to 10 upon the event $E_c$(C,S) originated in the client process, which is greater by 1 than the current logical clock of the client process. Later, the client logical clock is updated to 14 upon $E_c$(S,C). The internal events are time-stamped
by these logical clocks while communication events are not.

\setlength{\tabcolsep}{11.5pt}
\begin{table}[t]
  \centering
  \caption{\textsc{A Full Method-execution Event Sequence of the Example Program $E$.}}
  \vspace{-6pt}
    \begin{tabular}{|l|r||l|r|}
    \hline
    \multicolumn{2}{|l||}{\textbf{Server process}} & \multicolumn{2}{l|}{\textbf{Client process}} \\
    \hline
    Method Event & Timestamp & Method Event & Timestamp \\
    \hline
    {\tt S::main}$_e$ & 0 & {\tt C::main}$_e$ & 0 \\
    \hline
    {\tt S::init}$_e$ & 1 & {\tt C::init}$_e$ & 1 \\
    \hline
    {\tt S::init}$_i$ & 2 & {\tt C::init}$_i$ & 2 \\
    \hline
    {\tt S::init}$_x$ & 3 & {\tt C::init}$_x$ & 3 \\
    \hline
    {\tt S::main}$_i$ & 4 & {\tt C::main}$_i$ & 4 \\
    \hline
    {\tt S::serve}$_e$ & 5 & {\tt C::compute}$_e$ & 5 \\
    \hline
    $E_c$(C,S) & - & {\tt C::shuffle}$_e$ & 6 \\
    \hline
    {\tt S::getMax}$_e$ & \textbf{10} & {\tt C::shuffle}$_i$ & 7 \\
    \hline
    {\tt S::getMax}$_i$ & 11 & {\tt C::shuffle}$_x$ & 8 \\
    \hline
    {\tt S::getMax}$_x$ & 12 & {\tt C::compute}$_i$ & 9 \\
    \hline
    {\tt S::serve}$_i$ & 13 & $E_c$(C,S) & - \\
    \hline
    $E_c$(S,C) & - & $E_c$(S,C) & - \\
    \hline
    {\tt S::serve}$_x$ & 14 & {\tt C::compute}$_x$ & \textbf{14} \\
    \hline
    {\tt S::main}$_i$ & 15 & {\tt C::main}$_i$ & 15 \\
    \hline
    {\tt S::main}$_x$ & 16 & {\tt C::main}$_x$ & 16 \\
    \hline
    \end{tabular}%
  \label{tab:examplelts}%
  \vspace{-16pt}
\end{table}%

Next, suppose the query set $M$=$\{${\tt S::getMax}$\}$, {\distea} merges event traces of the two processes and,
by inferring impact relations from the timestamped events, it gives
$\{${\tt S::getMax}, {\tt S::serve}, {\tt S::main}, {\tt C::compute}, {\tt C::main}$\}$ as the impact set of $M$.
As is demonstrated, {\distea} can predict impacts across distributed components (processes). 
For instance, if the developer plans for a change to method {\tt getMax} in the server code, the methods {\tt compute} and {\tt main} in the client code, in addition to the other two server methods, are potentially affected and, thus, need inspections by the developer before applying that change.

\subsection{Analysis Algorithms}
\subsubsection{Partial Ordering of Internal Events}\label{subsec:ltsalgo}
%
Preserving the partial ordering of internal events is at the core of {\distea}, for which
two main options exist, both based on 
a logical notion of time: the {\LTS} approach~\cite{lamport1978time} as described before, and vector clocks~\cite{fidge1988timestamps,mattern1989virtual}.
In comparison, {\LTS} is lighter-weight as it just maintains a single counter as the logical clock for each process,
while a vector clock keeps an array of clocks for all processes. 
Therefore, we adopt {\LTS} in this work since it suffices for the current {\distea} design. 

%
%
Algorithm~\ref{algo} summarizes in pseudo code the {\distea} algorithm for partially ordering internal events based on
the original {\LTS}. 
The logical clock of the current process $C$ is initialized to 0 upon process start, as is the global
variable {\tt remaining}, 
which tracks the remaining length of data most recently
sent by the \emph{sender} process. The rest of this algorithm consists of two parts, which are
trigged upon the occurrence of communication events during system executions.

The first part is the runtime monitor \Call{SendMessage}{} trigged online upon each message-sending event.
The monitor piggybacks (prepends) two extra data items to the original message: the total length $sz$ of the data to send,
and the present value of the local logical clock $C$ (of this \emph{sender} process) (lines 2--3); then, it
sends out the packed data (line 4).

The second part is the other monitor \Call{RecvMessage}{}, which is trigged online upon each message-receiving event.
After reading the incoming message into a local buffer $d$ (line 6), the monitor decides whether to simply update the
size of remaining data and return (lines 7--9),
or to extract two more items of data first: the new total data length to read,
and the logical clock of the peer \emph{sender} process (lines 10--16).
In the latter case, the two items are retrieved, and then removed also, from the entire incoming message (lines 10--11).
Next, the remaining data length is reduced by the length of data already read in this event, and
 the local logical clock (of this \emph{receiver} process) is compared to the received one, updated to the greater, and incremented by 1 (lines 13--15). Lastly, the monitor returns the message as originally sent in the system
 (i.e., with the prepended data taken away).

To avoid interfering the message-passing semantics of the original system, {\distea}
keeps the length of remaining data (with the variable {\tt remaining} in the algorithm)
to determine the right timing for logical-clock retrieval.
In real-world distributed programs (e.g., Zookeeper~\cite{zookeeper}), 
it is common that 
a \emph{receiver}
process may obtain, through several reads, the entire data sent in a single write by its peer \emph{sender} process.
For example, a first read just retrieves data length so that an appropriate size of memory can be allocated to
take the actual data content in a second read.
Therefore, not only is it unnecessary to attempt retrieving
the prepended data items (data length and logical clock) in the second read
since the first one should have already done so, but also such attempts can break the original network I/O protocols.

\setlength{\textfloatsep}{0pt}
\begin{algorithm}[t]
\scriptsize{
    \caption{Monitoring communication events}
    \footnotesize{let $C$ be the logical clock of the current process} \\
    \footnotesize{remaining = 0 // remained length of data to read}
    \label{algo}
    \begin{algorithmic}[1]
        \Function{SendMessage}{$msg$} // on sending a message $msg$
            \State $sz$ = length of $sz$ + length of $C$ + length of $msg$
            \State pack $sz$, $C$, and $msg$, in order, to $d$
            \State write $d$
        \EndFunction
        \Function{RecvMessage}{$msg$} // on receiving a message $msg$
            \State read data of length $l$ into $d$ from $msg$
            \If{remaining $>$ 0}
                \State remaining -= $l$
                \State\Return $d$
            \EndIf
            \State retrieve and remove data length $k$ from $d$
            \State retrieve and remove logical clock $ts$ from $d$
            \State remaining = $k$ - length of $k$ - length of $ts$ - $l$
            \If{$ts$ $>$ $C$} 
                \State $C$ = $ts$
            \EndIf
            \State increment $C$ by 1
            \State\Return $d$
        \EndFunction
    \end{algorithmic}
}
\end{algorithm}

\subsubsection{Monitoring Internal Events}\label{subsec:easalgo}
Impact inference in {\distea} relies on the execution order of methods that is
deduced from the timestamps attached to all internal events, for which
{\distea} monitors the occurrence of each internal event.
However, 
as proved in~\cite{apiwattanapong05may}, recording just the \emph{first} entrance and \emph{last} returned-into
 (or return) events is equivalent to tracing the full sequence of those events for
the dynamic impact analysis in {\EAS}. Similarly, this equivalence also applies in {\distea}.
Thus, instead of keeping the timestamp for every internal-event occurrence (as shown in Table~\ref{tab:examplelts}),
{\distea} only 
records two key timestamps for each method $m$: 
the one for the first instance of $m_e$, and the one for the last instance of $m_i$ or $m_x$,
whichever occurred later. 

Accordingly, the online algorithm for monitoring internal events uses two counters 
to record the two key timestamps for each method, similar to what {\EAS} did but different in
that it does so \emph{in each process}.
Also, we use the per-process logical clock, instead of a global integer as used by {\EAS},
to update the per-method counters during runtime.
In the meanwhile, the logical clock $C_i$ of each process $P_i$ is maintained as follows:
\begin{itemize}[leftmargin=12pt]
\denseitems
\item Initialize $C_i$ to 0 upon the start of $P_i$.
\item Increase $C_i$ by 1 upon each internal event occurred in $P_i$.
\item Update $C_i$ upon each communication event occurred in $P_i$ via the two online monitors shown in Algorithm~\ref{algo}.
\end{itemize}
Finally, for the offline impact computation in {\distea}, the online algorithm here also dumps per-process internal-event sequences (i.e., the two timestamps for each executed method) as traces upon program termination.

\subsubsection{Impact Computation}
%
During system executions, the online internal-event monitoring algorithm
generates event traces concurrently and commonly on distributed machines.
Since it computes impacts offline, {\distea} gathers these traces to
one machine before merging them.
%
%
Then, from the merged traces, {\distea} computes the impact set of any given query
by searching methods that have the execute-after relation with that query according to
Equation~\ref{eq_is}.



\section{Implementation}
 \label{sec:impl}
The {\distea} tool\footnote{\vspace{-6pt}Download of the entire package is available at \href{http://nd.edu/~hcai/distea}{\url{http://nd.edu/~hcai/distea}.}}
consists of three main modules: a static analyzer, two sets of runtime monitors, and a post-processor.

\subsection{Static Analyzer}
The static analyzer instruments the input program such that
all method-execution events are monitored accurately, which is crucial to the soundness and precision of {\distea}.
We used Soot~\cite{lam11oct} 
for the instrumentation in two main steps.
%
First, {\distea} inserts probes for the three types of internal events in each method, for which we reused relevant
modules of {\diver}~\cite{cai14diver}.
%
The second step is to insert probes for communication events, for which {\distea} uses the
list $L$ of message-passing APIs, if specified, to identify probe points based on string matching:
$L$ includes the 
prototype of each unique API used in the input system for network I/Os.
%
If $L$ is not specified, a list of basic Java network I/O APIs will be used
covering two common means of blocking and non-blocking communications
: Java Socket I/O~\cite{javasocket} and Java NIO~\cite{javanio}.


\subsection{Runtime Monitors}
The two sets of runtime monitors implement the two online algorithms: the first for monitoring internal events
and the second preserving the partial ordering of them.
The first set again reuses relevant parts of {\diver}~\cite{cai14diver}.
For the second set, instead of invoking additional network I/O API calls to transfer logical clocks,
the monitors take over the original message passing
so that they can 
\emph{piggyback} the two extra data items (i.e., the data length and logical clock) to the original message. To that end,
the probes for the monitors \emph{replace} the original network I/O API calls during the instrumentation.

That is,
the extra data items are carried on by the original message passing. Our experience
suggested that this piggyback strategy is more viable than inserting additional calls, 
especially when dealing with selector-based non-blocking communications~\cite{artho2013software}.
For instance, the ShiVector tool in~\cite{abrahamson2014shedding}, which adopted the latter,
was unable to work with two of our subject programs (NioEcho and ZooKeeper).
One reason as we found 
is that, for a pair of an original call and the corresponding additional call, the two messages may not be read in
the same order by the \emph{receiver} process as in which they are sent by the \emph{sender} process. As a result,
an original message-receiving call may encounter extraneous data in the message hence the violation of original
network I/O semantics.

\subsection{Post-processor}
The post-processer is the module that actually answers impact-set queries. To that end, it
collects distributed traces through a helper script which passes per-process traces to
the offline impact-computation algorithm.
To compute impact sets, the post-processor retrieves the partial ordering of internal events by
just comparing their timestamps.

\section{Empirical Evaluation}
 \label{sec:eval}
To evaluate our approach, we conducted an empirical study to answer
the following
three research questions:

\begin{itemize}[leftmargin=14pt]
\denseitems
\item \textbf{RQ1} How effective is {\tech} in predicting impacts relative to existing alternative options?
%
\item \textbf{RQ2} How does impacts within processes compare with impacts across process boundaries?
%
\item \textbf{RQ3} How efficient and scalable is {\tech} in terms of the time and storage overheads it incurs?
\end{itemize}

The main goal of this evaluation was to investigate the effectiveness (RQ1) and
efficiency (RQ3) of {\distea}.
We also intended to examine the composition of {\distea} impact sets 
concerning how impacts propagate within and across distributed components (processes)
(RQ2).

\subsection{Experiment Setup}
We evaluated {\distea} on four distributed Java programs, as summarized in Table~\ref{tab:subjects}.
The size of each subject is measured by the number of
non-comment non-blank source lines of code (\emph{\#SLOC}), and number of
methods defined in the subject, both in Java, that we actually analyzed.
The last two columns list the input sets we used in our study, 
including the type and size of each set, and the number of methods (\emph{\#Queries})
covered in that set that we all used as impact-set queries.

MultiChat~\cite{multichat} is a chat application where multiple clients exchange
messages via a 
server broadcasting the message sent by one client to all others.
NioEcho~\cite{nioecho} is an 
echo service 
via which the client just gets back the same message as it sends to the server.
ZooKeeper~\cite{zookeeper,hunt2010zookeeper} is a 
coordination service for distributed systems
to 
achieve consistency and synchronization.
Voldemort~\cite{voldemort} is a distributed key-value storage system used at LinkedIn.
The first two use only Socket I/O and Java NIO, respectively, and the last two used both.
For all subjects, we checked out from their official online repositories for
the latest versions or revisions as shown in (the parentheses of) Table~\ref{tab:subjects}.

\setlength{\tabcolsep}{5.5pt}
\begin{table}[t]
  \centering
  \vspace{-4pt}
  \caption{\textsc{Statistics of Experimental Subjects}}
    \vspace{-8pt}
    \begin{tabular}{|l||r|r||r|r|}
    \hline
    \textbf{Subject} & \textbf{\#SLOC} & \textbf{\#Methods} & \textbf{Inputs (size and type)} & \textbf{\#Queries} \\
    \hline
    \multicolumn{1}{|l||}{\shortstack{MultiChat (r5)}} & 470 & 37 & 1 integration  test & 25 \\
    \hline
    \multicolumn{1}{|l||}{\shortstack{NioEcho (r69)}} & 412 & 27 & 1 integration  test & 26 \\
    \hline
    \multicolumn{1}{|l||}{\multirow{4}[5]{*}{\shortstack{ZooKeeper\\(v3.4.6)}}} & \multirow{4}[5]{*}{62,450} & \multirow{4}[5]{*}{4,813} & 1 integration  test & 749 \\
\cline{4-5}    \multicolumn{1}{|l||}{} &   &   & 1 system test & 817 \\
\cline{4-5}    \multicolumn{1}{|l||}{} &   &   & 1 load test & 798 \\
\cline{4-5}    \multicolumn{1}{|l||}{} &   &   & 195 unit tests & 2,780 \\
    \hline
    \multicolumn{1}{|l||}{\multirow{4}[5]{*}{\shortstack{Voldemort\\(v1.9.6)}}} & \multirow{4}[5]{*}{163,601} & \multirow{4}[5]{*}{17,843} & 1 integration  test & 2,048 \\
\cline{4-5}    \multicolumn{1}{|l||}{} &   &   & 1 system test & 1,056 \\
\cline{4-5}    \multicolumn{1}{|l||}{} &   &   & 1 load test & 1,323 \\
\cline{4-5}    \multicolumn{1}{|l||}{} &   &   & 9 unit tests & 3,421 \\
    \hline
    \end{tabular}%
  \label{tab:subjects}
  \vspace{0pt}
\end{table}%

We chose these subjects 
such that a variety of program sizes, application domains, and uses of both blocking and non-blocking I/Os are all considered;
we chose the input sets to cover different types of inputs when possible, including system test, integration test, load test, and unit test. Except for the integration tests, other types of inputs come
with the subjects as integral parts. For unit tests, we used only those leading to multiprocess executions from
the full original sets.

For each subject, we created the single integration test in which we started one server process
and one client process on two separated machines and manually performed client operations that
cover basic server functionalities.
Specifically, for MultiChat and NioEcho, the client requests were sending random text messages;
%
%
%
for ZooKeeper, the client operations were, in order: create a node,
look up for it, check its attributes, change its data association, and
delete it;
for Voldemort, the operations were, also in order: add a key-value pair,
query the key for its value, delete the key, and retrieve the pair again.
%

\subsection{Experimental Methodology}\label{sec:methodology}
%
%
To the best of our knowledge, there are no other dynamic impact analysis techniques for distributed systems in the literature.
We could not compare to slicing techniques either as we are not aware of 
such slicers readily available to us and working with real-world distributed systems like our subjects,
while developing one 
would require considerable efforts.

Therefore, we assume two possible alternatives to {\distea}: ignoring impacts outside the process where the query first executed (referred to as \emph{local process}, versus all others as \emph{remote processes});
taking all methods executed in remote processes as impacted. In contrast, the latter (i.e., method-level coverage, referred to as \emph{MCov}) is safe hence potentially a more practical option in most cases.
In fact, \emph{MCov} is an easy adaptation for distributed systems from {\CI}~\cite{orso03sep}, a major existing option for centralized programs.
Thus, we consider \emph{MCov} as the baseline technique and
the covered (executed) sets of methods as baseline impact sets. 
We refer to impacts in local and remote processes as \emph{local impacts} and \emph{remote impacts}, respectively.

For every method of each subject as the query,
we measure the effectiveness of {\distea} by comparing the impacts it predicted to those given by \emph{MCov} in terms
of 
impact-set size ratios, and
examine the composition of the impact set concerning its two
subsets: \emph{local impact set} and \emph{remote impact set}, while also analyzing their intersection, referred to as \emph{common impact set}.
Accordingly, we measure too the effectiveness of {\distea} with respect to these subsets relative to the corresponding
\emph{MCov} results.

When computing the impact set for a query, for each type of inputs, we take the union of per-input impact sets of all
inputs of that type 
since each type 
is usually intended to represent a different operational profile of the system under analysis.
Finally, beside the impact querying time, we report the static-analysis and runtime costs of {\distea}, and
storage costs of event traces, together as efficiency metrics. All machines used in our experiments are Linux workstations
with an Intel i5-2400 3.10GHz processor and 8GB DDR2 RAM.


\subsection{Results and Analysis}
\subsubsection{RQ1: Effectiveness}
\begin{figure*}[t]
  \vspace{-4pt}
  \centering
  \includegraphics[scale=0.8]{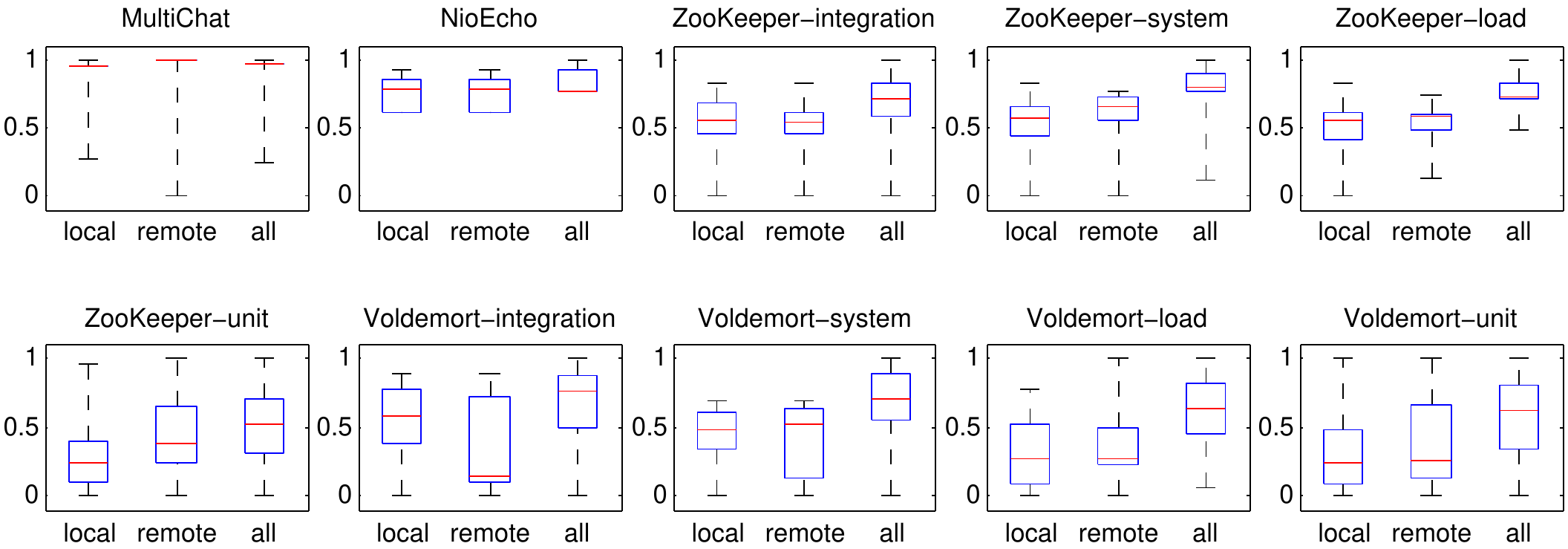}
  \vspace{-6pt}
  \caption{Effectiveness of {\distea} expressed as the ratios ($y$ axes) of its per-query impact-set sizes, including those of the local and remote subsets ($x$ axes), versus \emph{MCov} as the baseline, for each subject and input-set type (atop each plot as the title).}
  \label{fig:eff}
  \vspace{-24pt}
\end{figure*}
Figure~\ref{fig:eff} shows the main effectiveness results, with one plot depicting the data distribution
for each subject and input type, shown as the plot title (hereafter, the input type is omitted for
the two small subjects as only integration test is used for them).
Each plot includes three box plots showing that data distribution for one of three categories (on $x$ axis):
the holistic impact set (\emph{all}) and its two subsets (\emph{local} and \emph{remote}),
with each data point indicating the effectiveness metric (on $y$ axis) for one query.

The results show that {\distea} is constantly more effective than \emph{MCov}, reducing the impact sets of the latter
by 15\% to over 95\%.
Compared to the small subjects, the two large subjects see noticeably better effectiveness, possibly because
the few methods in the small ones tend to all focus on
a single task hence 
have more methods executed after any other methods.
Also, the ratios with respect to \emph{all} impact sets are always higher than those to the two subsets.
The reason is that
the two subsets from both approaches have substantial intersections, 
while
the ones from {\distea} are consistently smaller than those from \emph{MCov}. 


Overall, as shown by the complementary results, the mean effectiveness, in Table~\ref{tab:costs} (left four columns),
{\distea} reports on average only 64\% of the impacts produced by \emph{MCov}.
In particular, the reductions in remote impact sets are even higher, by 56\% on average and well above 50\% in
most individual cases. This implies that, relative to the baseline, 
developers can save the time that would
be spent on inspecting more than half of the impacts propagated in remote processes.
\subsubsection{RQ2: Impact-set Composition}
\begin{figure*}[t]
  \vspace{-4pt}
  \centering
  \includegraphics[scale=0.8]{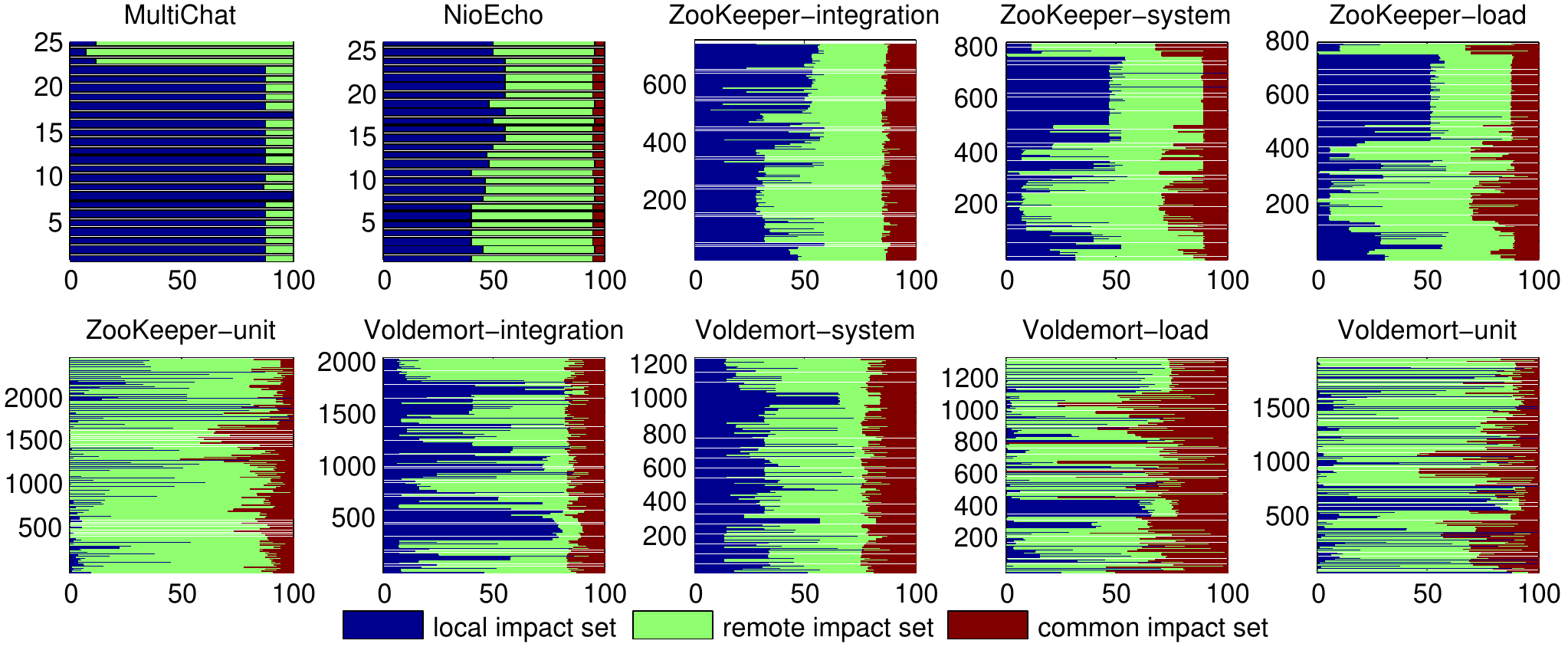}
  \vspace{-4pt}
  \caption{Composition of impact sets given by {\distea} for all queries ($y$ axes) expressed as the percentages ($x$ axes) of local, remote, and common impact sets in the whole impact set per query, for each subject and input-set type (atop each plot).}
  \label{fig:comp}
  \vspace{-20pt}
\end{figure*}

Figure~\ref{fig:comp} plots the impact-set composition for each individual
query numbered on the $y$ axis, where the $x$ axis indicates the percentage of three complementary sets, \emph{local}, \emph{remote}, and \emph{common}, 
for each subject and input-set type.
The common sets have been removed from the local and remote subsets, but in this figure only, to help clarify the composition.

A first observation is that 
remote impact sets 
dominate corresponding
holistic impact sets in way most cases. 
%
For one thing, this might explain the mostly higher impact-set size ratios for remote impacts than for local ones,
as seen in Figure~\ref{fig:eff}.
For another, the contrast in size between the two subsets
suggests that impacts can propagate much more
largely to remote processes than in local ones in distributed programs.

Another finding is that, for almost all queries, there were methods executed after the query in both 
local and remote processes. 
This implies that in distributed systems, components often
share 
common functionalities. Moreover, the sizes of common impact sets could be 
a metric of functional overlapping and code reuse 
among components of distributed systems. 
Also, the figure shows that the strength of this metric seems to continuously increase with the system size.

\subsubsection{RQ3: Efficiency}
\setlength{\tabcolsep}{8pt}
\begin{table*}[t]
  \centering
  \caption{\textsc{Mean Effectiveness, Time-Cost Breakdown, and Storage Costs of {\distea}}}
  \vspace{-8pt}
    \begin{tabular}{|l||r|r|r||r|r|r|r|r||r|}
    \hline
    \multirow{2}[4]{*}{\textbf{Subject \& input}} & \multicolumn{3}{c||}{\textbf{Mean impact-set size ratios}} & \multicolumn{5}{c||}{\textbf{Time costs (ms)}} & \multicolumn{1}{c|}{\multirow{2}[4]{*}{\textbf{\shortstack{Storage\\costs (KB)}}}} \\
\cline{2-9}      & \multicolumn{1}{c|}{\textbf{Local}} & \multicolumn{1}{c|}{\textbf{Remote}} & \multicolumn{1}{c||}{\textbf{All}} & \multicolumn{1}{c|}{\textbf{\shortstack{Static\\analysis}}} & \multicolumn{1}{c|}{\textbf{\shortstack{Normal\\run}}} & \multicolumn{1}{c|}{\textbf{\shortstack{Instrumented\\run}}} & \multicolumn{1}{c|}{\textbf{\shortstack{Runtime\\overhead}}} & \multicolumn{1}{c||}{\textbf{\shortstack{Querying\\(with stdev)}}} & \multicolumn{1}{c|}{} \\
    \hline
    MultiChat & 85.33\% & 85.03\% & 86.08\% & 12,817 & 5,461 & 5,738 & 5.07\% & 3 (2) & 6 \\
    \hline
    NioEcho & 77.07\% & 76.42\% & 83.73\% & 13,365 & 3,213 & 3,623 & 12.76\% & 3 (3) & 4 \\
    \hline
    ZooKeeper-integration & 55.85\% & 53.83\% & 70.73\% & \multirow{4}[8]{*}{39,124} & 37,239 & 38,416 & 3.16\% & 10 (3) & 96 \\
\cline{1-4}\cline{6-10}    ZooKeeper-system & 53.54\% & 62.22\% & 81.17\% &   & 15,385 & 18,578 & 20.75\% & 16 (8) & 136 \\
\cline{1-4}\cline{6-10}    ZooKeeper-load & 50.14\% & 55.07\% & 76.17\% &   & 94,187 & 98,930 & 5.04\% & 15 (7) & 140 \\
\cline{1-4}\cline{6-10}    ZooKeeper-unit & 28.56\% & 43.26\% & 51.44\% &   & 1,109,146 & 1,370,143 & 23.53\% & 14,619 (87,056) & 188,804 \\
    \hline
    Voldemort-integration & 55.86\% & 40.12\% & 69.72\% & \multirow{4}[8]{*}{132,536} & 17,755 & 18,697 & 5.31\% & 27 (9) & 312 \\
\cline{1-4}\cline{6-10}    Voldemort-system & 45.71\% & 42.36\% & 67.12\% &   & 11,136 & 12,253 & 10.03\% & 19 (7) & 196 \\
\cline{1-4}\cline{6-10}    Voldemort-load & 30.90\% & 37.50\% & 61.68\% &   & 21,066 & 21,253 & 0.89\% & 122 (190) & 776 \\
\cline{1-4}\cline{6-10}    Voldemort-unit & 31.77\% & 37.55\% & 57.50\% &   & 132,676 & 167,861 & 26.52\% & 403 (835) & 6,984 \\
    \hline
    \textbf{Overall average} & \textbf{41.42\%} & \textbf{43.99\%} & \textbf{63.88\%} & \textbf{49,460.5} & \textbf{144,726.4} & \textbf{175,549.2} & \textbf{11.31\%} & \textbf{3,228.9 (40,752.1)} & \textbf{19,745.4} \\
    \hline
    \end{tabular}%
  \label{tab:costs}%
  \vspace{-12pt}
\end{table*}%
Table~\ref{tab:costs} 
lists all relevant costs of {\distea} for this study,
including the runtime of static analysis,
runtime overhead measured as ratios of the runtime of the original
program (\emph{Normal run}) over the instrumented one (\emph{Instrumented run}), and impact querying time.
%

The static 
analysis took longer for larger subjects as expected, yet still within 2.2 minutes even
on the largest system Voldemort. Note that this is a one-time cost (for the single program version analyzed by {\distea} at least) as the instrumented code can be executed on any inputs and used for computing any queries afterwards. 
%
Runtime and querying costs are constantly correlated to subject sizes as well as the sizes of
input sets, with the worst case seen by ZooKeeper on its unit-test input set, which is by far the largest
among all subjects and inputs studied.
Nevertheless, the runtime overhead is at worst 26\% and the longest querying time is in 
15 seconds.

Storage costs are also tightly connected to the number of inputs in addition to subject sizes, of
which the largest is 188MB 
for the 195 traces of ZooKeeper. In other cases, 
this cost is at most 7MB, with an overall average less than 20MB.

In all, the results suggest that {\distea} is highly efficient in both time and space dimensions, and that it
seems to be readily scalable up to large systems.
As shown in the bottom row of the table, it costs on average less than one minute for static analysis and a couple
seconds for computing the impact set per query, with the mean runtime overhead of about 10\%.

\subsection{Threats to Validity}
The main threat to \emph{internal} validity lies
in possible implementation errors in {\distea} and experiment scripts.
To reduce this threat, we did a careful code review for our tools and
used the two small subjects 
to manually validate
their functionalities and analysis results.
An additional such threat concerns about possible missing (remote) impacts due to
network I/Os that were not monitored at runtime. However, we checked the code
of all subjects and confirmed that they only used the most common message-passing APIs monitored
by our tool with respect to their input sets that we studied.


The main threat to \emph{external} validity is that our study results may not generalize to
other distributed programs and input sets. In this study, we considered only limited
number of subjects, which may not represent all real-world programs, and only subsets of inputs,
which do not necessarily represent all behaviours of the studied programs.
To reduce this threat, we have chosen programs of various sizes and application domains, including
the two enterprise systems in different areas. In addition, we considered
different types of inputs, including integration, system, load, and unit tests.
Most of these tests came as part of the subjects except for the integration tests, which
we created according to the official online design documentation of these programs.

The main threat to \emph{construct} validity is the metrics used for the evaluation.
Without directly comparable peer techniques in the literature,
we assumed that developers would use coverage-based approach like \emph{MCov}, as a representative
alternative to {\distea}, to narrow down the search space of potential impacts in
the context of distributed 
executions.
To mitigate this threat, we examined the composition of each impact set and analyzed
the effectiveness with respect to its local and remote subsets in addition to that
of the holistic impact set to help demonstrate the usefulness of {\distea}.

%

Finally, a \emph{conclusion} threat concerns about the data points analyzed:
We applied the statistical analyses only to methods for which
impact sets could be queried (i.e., methods executed at least once).
Also, the present study only considered potential changes in single methods for each query, while in practice
developers may plan for changes in multiple methods at a time, which may lead to different 
results.
To minimize this threat, we adopted this strategy for all experiments and calculated
the experimental metrics for every possible query.


\section{Case Studies}\label{sec:appendix}
To further investigate the effectiveness and, more important, the practical utility of {\tech}, we conducted two case studies.
In contrast to the foregoing empirical evaluation, the case studies were focused on sample subjects and inputs against
a small number of queries for in-depth examination.
%

\subsection{Study I: Precision and Usefulness of {\distea} Impact Sets}
\subsubsection{Methodology}
As we discussed earlier, {\distea} can be imprecise. 
Yet, currently there is no automatic means available for us
to assess exactly how imprecise it would be.
Thus, in our first case study, we investigate the precision of {\distea}.
To that end, we randomly chose two queries from a small subject MultiChat and three from a large one Voldemort, and
picked an input set for each also randomly.
We then manually determine the ground-truth impact set 
of the chosen queries 
according to our understanding
of the system's runtime behaviour with respect to the input.
Since the manual inspection 
is exhaustive, we limited our choices of
queries and inputs to those for which the {\distea} impact sets had no more than 50 methods.

\subsubsection{Results}
Table~\ref{tab:casestudy1} lists the results for the five cases we studied.
For each case, the table summarizes the impact-set sizes (\emph{IS})
from {\tech}, manual inspection, and the baseline approach (\emph{MCov}), all separately for the two subsets
(\emph{local} and \emph{remote}).
The numbers in parentheses are the precision 
of {\distea}
(recall was constantly 100\%).\footnote{\vspace{-6pt}Details on the results are at  \href{http://nd.edu/~hcai/distea/casestudy1.html}{\url{http://nd.edu/~hcai/distea/casestudy1.html}}.}

\setlength{\tabcolsep}{2.5pt}
\begin{table}[htbp]
  \centering
  \caption{\textsc{Results for Five Cases of Using {\tech}}}
  \vspace{-8pt}
    \begin{tabular}{|r|r|r|r|r|r|r|}
    \hline
    \multicolumn{1}{|c|}{\multirow{2}[4]{*}{\textbf{Subject \& input}}} & \multicolumn{2}{c|}{\textbf{{\tech} IS (precision)}} & \multicolumn{2}{c|}{\textbf{Manual true IS}} & \multicolumn{2}{c|}{\textbf{\emph{MCov} IS}} \\
\cline{2-7}    \multicolumn{1}{|c|}{} & \multicolumn{1}{c|}{local} & \multicolumn{1}{c|}{remote} & \multicolumn{1}{c|}{local} & \multicolumn{1}{c|}{remote} & \multicolumn{1}{c|}{local} & \multicolumn{1}{c|}{remote} \\
    \hline
    \multicolumn{1}{|l|}{MultiChat} & 1 (100\%) & 13 (69.2\%) & 1 & 9 & 3 & 21 \\
    \hline
    \multicolumn{1}{|l|}{MultiChat} & 13 (76.9\%) & 2
(50\%) & 10 & 1 & 22 & 3 \\
    \hline
    \multicolumn{1}{|l|}{Voldemort-system} & 4 (100\%) & 23 (56.5\%) & 4 & 13 & 740 & 809 \\
    \hline
    \multicolumn{1}{|l|}{Voldemort-system} & 3 (33.3\%) & 0 (-) & 1 & 0 & 811 & 440 \\
    \hline
    \multicolumn{1}{|l|}{Voldemort-load} & 13 (46.1\%) & 41 (41.4\%) & 6 & 17 & 288 & 500 \\
    \hline
    \multicolumn{1}{|l|}{\textbf{Overall average}} & 6.8 (\textbf{71.2\%}) & 15.8 (\textbf{51.7\%}) & 4.7 & 8 & 373 & 354.6 \\
    \hline
    \end{tabular}%
  \label{tab:casestudy1}%
\end{table}%

In most of these cases, a considerable portion of the impact sets was methods executed after but not
to be impacted by the query, as we expected. For instance, the first query in MultiChat was the {\tt run} method of
the main client thread, which executed at the end of the client process to iteratively send user inputs to the server.
As a result, in the local process the query could only impact itself; 
and of 13 methods executed after it in
the server process, four were false positives as they just dealt with network connections
independently of any specific message received from clients.

For another example, the last query in Voldemort, an error-handling utility method, is defined in a common module executed by both the Voldemort master and its clients. Among six common impacts reported by {\tech}, only two were possibly to be impacted by the query. The other four, along with three of the seven unique impacts in the local process, never involved error handling. These methods, as 24 out of 41 methods devoted to network service maintenance only in the remote process, were falsely reported
as impacted.

In all, {\tech} had an overall average precision of 56.9\% for the five randomly selected queries, and
for remote impact sets only the number was 51.7\%. These are very close to
the precision of {\EAS} obtained from an extensive study using various types of changes in~\cite{cai15jss}.
In contrast,
precision of the local impact sets was considerably higher, not only on overall average but constantly in every single case.
This may be due to the even looser coupling, via message passing only, among methods across components than within components.

Since we only studied five cases, these results are by no means conclusive. Yet, it seems to suggest that developers
could expect impact sets close to 60\% precise from {\tech} in an average case. Note that although this precision may not be
sufficient in some situations, such as when reported impact sets are extremely large, {\distea} is reasonably effective and
useful with respect to the much larger sets of covered methods (\emph{MCov} results).
For instance, 
checking on average 16 methods only with {\distea}, instead of 354 with \emph{MCov}, for impacts
propagated to remote components implies significant reduction in impact-inspection efforts for the studied cases.

\subsection{Study II: Utility for Distributed-Program Understanding}
\subsubsection{Methodology}
Dynamic analysis is an important means for program comprehension based on
concrete executions,
a process on which developers often spend a great deal of effort~\cite{cornelissen2009systematic}.
However, understanding distributed system executions is a challenging task because of the complex
interactions among concurrent component executions in such systems~\cite{moc2001understanding},
and even more so in the presence of selector-based non-blocking communications~\cite{artho2013software}.

Thus, our second case study aimed to explore
the utility of {\distea} for program comprehension of distributed systems.
In particular, we intended to see if {\distea} can help 
understand the interface between
distributed components and interprocess communications (IPC) among them.
For this study, we chose NioEcho and ZooKeeper, both of which utilize selector-based network I/Os
(Java NIO~\cite{artho2013software}) for IPC.
For each of these two subjects, without prior knowledge about internals of either, we
first picked a few important-looking queries (simply based on names) from each component, and
then executed the instrumented program on the same integration test as used in the empirical study.
Next, we took the {\distea} impact sets of the selected queries to learn about its
runtime IPC semantics.

\subsubsection{Results}
For NioEcho, despite of its small source size and simple high-level functionality,
the non-blocking IPCs between the server and client made
it much harder than expected to fully understand the program, interactions
between the two components in particular, by just reading its source code.
To use our tool, we picked two queries 
from the client and three from the server 
that all \emph{looked} closely relevant to messaging.
It turned out that using {\distea} was quite instrumental
in this case: the local and remote impacts, when listed together in the ascending order of
associated timestamps, clearly show how the client initiates a response handler for a message before
sending it out and, before it gets to wait for response in the handler by checking the selector it registered,
the server already received the message and started its echo service, after which the client reads the
server response.

In the case of ZooKeeper, given
the large size and complexity of the entire system, 
we targeted only the particular IPCs with respect to one fundamental operation {\tt getData}~\cite{hunt2010zookeeper}.
We started with the entry methods of both the server and client modules, and then by searching in their impact sets
we located one most relevant-looking query from the client. 
Next, examining local and remote impacts of that query 
let us identify the major transaction steps: The client first prepares a data request
using an external library and then forwards the request to a client thread which spawned another child thread
to actually connect to the server; next, when the client proceeds with some bookkeeping routines while waiting for
response, the execute-after sequence ({\tech} results) in the remote process shows that the server 
has accepted the request and initiated a
thread to access a database, and then started another thread to send the retrieved data back, followed by the client's taking
the response and processing the data.

In both cases, we also found the ordering of impacts (by their timestamps) fairly useful
for following component-level interactions step by
step, and that the common impacts, which reveal code reuse among components, also facilitate the comprehension process.
On the other hand, however, we noticed that navigating in the textual impact sets can be tedious when the results become large,
for which some effective visualizations~\cite{moc2001understanding,kunz1997poet} would be helpful.
One possible solution, for example, is to visualize the partially ordered impacts as 
an \emph{interprocess} call graph, which 
can complement or collaborate with other distributed-program comprehension approaches, such as space-time diagrams and
communicating finite state machines~\cite{beschastnikh2014inferring}.

In sum, while the results of this exploratory study may not generalize to other cases and systems, our experience
suggests that {\distea} can help users understand distributed programs and their executions. Note that the main source of this benefit lies in the remote impacts {\distea} produces, which tell about the interactivity among distributed components.

\section{Discussion}
The core technique of {\distea} is to partially order distributed method-execution
events to discover 
execute-after relations among methods in multiple concurrent processes.
And we have demonstrated that the technique 
can be an important step 
for effectively evolving distributed systems.
On the other hand, its conservative nature, while makes it safe modulo the concrete executions utilized, 
can also lead to false positives.
Nonetheless, {\distea} 
will be a practically useful option for developers since it provides rapid, although possibly rough, results~\cite{jackson2000software}. {\distea} is highly efficient, whereas a more accurate analysis would
need to trade efficiency for precision if remaining sound.
Also, developers actually need multiple levels of cost-effectiveness tradeoffs for impact analysis~\cite{cai15diverplus}.

A few other limitations exist.
First, the present 
tool might not immediately fully work for arbitrary distributed systems, because it now considers
only the two common cases of message-passing APIs by default and, even with the user-specified list of all such APIs, the
text matching by API prototypes {\distea} uses for locating instrumentation points may cause incomplete
communication-event monitoring. However, this is the limitation of our current implementation, not of the technique.
%
In addition, the instrumentation for monitoring method events and extra network traffics for exchanging logical clocks
may affect the performance of original systems~\cite{beschastnikh2014inferring}, although we expect such effects to be
minor in general according to our studies.

\section{Related Work}
Three main categories of previous work are related to ours: dynamic impact analysis, dependence analysis of
concurrent programs, and logging and timing for distributed systems.
\subsection{Dynamic Impact Analysis}
The execute-after-sequences ({\EAS}) approach~\cite{apiwattanapong05may} which partially inspired {\distea} 
is a performance optimization of its predecessor {\PI}
~\cite{law03may}.
Many other dynamic impact analysis techniques also exist~\cite{Li2013ASC}, aiming at
improving precision~\cite{Breech2006IIM,Huang2007PDI,Hattori2008OTP,cai14diver}, recall~\cite{maia2010hybrid},
efficiency~\cite{Breech2004OIA,Breech2005ACO}, and cost-effectiveness~\cite{cai15diverplus} over {\PI} and {\EAS}.
However, these techniques did not address distributed or multiprocess programs that we focus on in this work.

Two recent advances in dynamic impact analysis,
{\diver}~\cite{cai14diver} and the multivariate framework in~\cite{cai15diverplus},
utilize hybrid program analysis to achieve higher precision and more flexible cost-effectiveness options over {\EAS}-based
approaches, but still focus on centralized programs.
As a first step, {\distea} sacrifices imprecision for high efficiency. However, it would be interesting to 
adopt hybrid approaches for distributed systems too. For instance, among other improvements, one may be to immediately gain
better analysis precision by first using static dependencies to prune false-positive impacts
\emph{within} each process, as {\diver} did, and then propagating impacts across process boundaries by means of the
Lamport timestamps as used in this work. More aggressive pruning may also be promising if leveraging
more and/or finer interprocess dependencies, such as communication dependencies and synchronization dependencies~\cite{kamkar1995dynamic,mohapatra2006distributed}, as exploited by distributed-program slicing techniques~\cite{barpanda2011dynamic}.

\subsection{Dependence Analysis of Concurrent Programs}
Using fine-grained 
dependency analysis,
a large body of work attempted to extend traditional slicing algorithms to concurrent
programs~\cite{krinke2003context,xiao2005improved,Nanda2006ISM,giffhorn2009precise,xu2005brief} yet
mostly focusing on centralized, and primarily multithreaded, ones.
For those programs, traditional dependence analysis was extended to handle additional
dependencies due to shared variable accesses, synchronization, and communication
between threads and/or processes (e.g.,~\cite{krinke2003context,Nanda2006ISM}).
While {\distea} also handles multithreaded programs, 
it 
targets 
multiprocess ones running on distributed machines, and aims at lightweight impact analysis instead of
fine-grained slicing.

For systems running in multiple processes where interprocess communications are realized
via socket-based message passing,
an approximation for static slicing was discussed in~\cite{krinke2003context}.
Various dynamic slicing algorithms have been proposed too, earlier for procedural programs only ~\cite{korel1992dynamic,cheng1997dependence,goswami2000dynamic, duesterwald1993distributed,kamkar1995dynamic} and
recently for object-oriented software also~\cite{mohapatra2006distributed,barpanda2011dynamic,pani2012slicing}.
And a more complete and detailed summary of slicing techniques for distributed programs can be found in~\cite{xu2005brief}
 and~\cite{barpanda2011dynamic}.
Although these slicing algorithms were rarely evaluated against large real-world distributed systems, it can be anticipated
that they would face scalability issues with large systems based on the limited
empirical results they reported and the heavyweight nature of their design.

In contrast to these fine-grained (statement-level) analysis,
{\distea} aims at a highly efficient
method-level dynamic impact analysis
that can readily scale up to large distributed programs.
A few more static analysis algorithms for distributed systems exist as well but focus on
 other (special) types of systems, such as RMI-based Java programs~\cite{sharp2006static} and Android applications~\cite{octeau2013effective}, different from the common type of distributed systems~\cite{Coulouris2011DSC} {\distea} addresses.





At coarser levels, other researchers resolve dependencies in 
distributed systems too but for different purposes such as 
enhancing parallelization~\cite{psarris2004experimental}, system configuration~\cite{kon2000dependence},
and high-level system 
modeling~\cite{abrahamson2014shedding,beschastnikh2014inferring}.
A static analysis, LSME
~\cite{murphy1996lightweight} extracts 
inter-component
dependencies due to implicit invocations, but it is both 
imprecise and unsound~\cite{popescu2012impact,garcia2013identifying}.
In~\cite{garcia2013identifying}, another static analysis is proposed to infer
inter-component dependencies based on messaging-interface matching.
In contrast,
{\distea} performs code-based analysis while providing more focused impacts relative to concrete program executions than
static-analysis approaches.

An impact analysis dedicated to distributed systems, Helios~\cite{popescu2012impact}
can predict impacts of potential changes to support evolution tasks for DEBS.
However, it relies on particular message-type filtering and manual annotations 
in addition to a few other constraints. Although these limitations are largely lifted
by its successor Eos~\cite{garcia2013identifying}, both approaches are 
\emph{static} and limited to DEBS only, as is the
 latest technique~\cite{Tragatschnig2014IAE} which identifies impacts based on change-type classification yet
 ignores intra-component dependencies hence provides merely incomplete results. 
While sharing similar goals, {\distea} targets a broader range of distributed systems than DEBS using \emph{dynamic} analysis and without relying on special source-code information (e.g., interface patterns) as those techniques do.
%

\subsection{Logging and Timing for Distributed Systems}
To facilitate high-level understanding of distributed systems and executions,
techniques like logging and mining runtime logs~\cite{beschastnikh2014inferring,lou2010mining}
infer inter-component interactions using textual analysis of system logs,
relying on the availability of particular data such as informative logs and certain patterns in them.
{\distea} also utilizes similar information (i.e., the Lamport timestamps) 
but infers the happens-before relations among method-execution events mainly for code-level impact analysis.
Also, {\distea} automatically generates such information it requires rather than
relying on existing information in the original programs 
(e.g., logging statements).

The Lamport timestamp
used by {\distea} is closely related to the vector clocks~\cite{fidge1988timestamps,mattern1989virtual} used by
other tools, such as ShiVector~\cite{abrahamson2014shedding} for ordering distributed logs and Poet~\cite{kunz1997poet}
for visualizing distributed systems executions.
While we could utilize vector clocks also, 
we chose the Lamport timestamp as it is lighter-weight
and simpler yet well suffices for this work. 
In addition, unlike ShiVector, which requires accesses to source code and recompilation using
the AspectJ compiler, {\distea} does not have such constraints as it works on bytecode. 

\section{Conclusion} 
Components in distributed systems usually run concurrently in separated processes and
communicate via socket-based message passing without explicitly invoking or
referencing each other. In consequence,
existing dynamic impact analysis, which relies on
explicit invocations (dependencies), tends to be either entirely inapplicable or
at best quite ineffective to use for those systems.

We presented {\distea}, a dynamic impact analysis for evolving common-type distributed systems.
By leveraging lightweight monitoring of method-execution events and partially ordering those events
during concurrent multiprocess executions, {\distea} can safely predict potential impacts
of any query both within and across all processes, supporting efficient dynamic impact analysis of
distributed systems. {\distea} has been implemented for Java and is publicly available for download.
%
Through an empirical evaluation and two case studies on four distributed Java programs,
including two large real-world enterprise systems, 
we have shown the high efficiency of {\distea} and its superior effectiveness over existing alternatives, and
also illustrated its benefits for understanding distributed systems and executions.
Overall, {\distea} offers a promising option to developers for maintaining and evolving distributed systems.

\newpage
\bibliographystyle{IEEEtran}
\balance
\bibliography{reference}



\end{document}